\documentclass[reprint,prx]{revtex4-2}

\usepackage[dvips]{color}
\usepackage{dcolumn}
\usepackage{bm}
\usepackage{psfrag}
\usepackage{amsmath,amssymb}
\usepackage{indentfirst}
\usepackage{graphicx}
\usepackage{hyperref}

\begin{document}
\title{Effect of N, C and B interstitials on the structural and magnetic properties of alloys with Cu$_3$Au-structure}
\author{Ingo Opahle$^{1}$}
\email{opahle@tmm.tu-darmstadt.de}
\author{Harish K. Singh$^{1}$}
\author{Jan Zemen$^{2,3}$}
\author{Oliver Gutfleisch$^{1}$}
\author{Hongbin Zhang$^{1}$}
\affiliation{$^{1}$ Institute of Materials Science, TU Darmstadt, 
64287 Darmstadt, Germany\\
$^{2}$ Faculty of Electrical Engineering, Czech Technical University in Prague, Technicka 2, Prague 166 27, Czech Republic\\
$^{3}$ Department of Physics, Blackett Laboratory, Imperial College London, London SW7 2AZ, United Kingdom}

\date{\today}
%

\begin{abstract}
High-throughput density functional calculations are used to investigate 
the effect of interstitial B, C and N atoms on 
21 alloys reported to crystallize in the cubic Cu$_3$Au structure.
It is shown that the interstitials can have a significant impact on the 
magneto-crystalline anisotropy energy (MAE), the thermodynamic stability 
and the magnetic ground state structure, making  these alloys
interesting for 
hard magnetic, magnetocaloric and other applications.
For 29 alloy/interstitial combinations the formation of stable alloys with 
interstitial concentrations above 5\% is expected.
In Ni$_3$Mn 
interstitial N induces a tetragonal distortion
with substantial uniaxial MAE for realistic N concentrations.
Mn$_3X$N$_x$ ($X$=Rh, Ir, Pt and Sb) are identified as alloys with strong
magneto-crystalline anisotropy. 
For Mn$_3$Ir  we find a strong enhancement of the MAE upon N alloying
in the most stable collinear ferrimagnetic state as well as in
the non-collinear magnetic ground state.
Mn$_3$Ir and Mn$_3$IrN show also interesting topological transport properties.
The effect of N concentration and strain 
on the magnetic properties are discussed.
Further, the huge impact of N on the MAE of Mn$_3$Ir and a possible impact 
of interstitial N on amorphous Mn$_3$Ir, a material that is indispensable in today's data storage devices,
are discussed at hand of the electronic 
structure.
For Mn$_3$Sb, non-collinear, ferrimagnetic and ferromagnetic states
  are very close in energy, making this material potentially interesting for
magnetocaloric applications.
For the investigated Mn alloys and competing phases, 
the determination of the magnetic ground state is essential for a
reliable prediction of the phase stability.
\end{abstract}  
\maketitle
\section{Introduction}
  Magnetic materials with a large magnetocrystalline anisotropy energy (MAE) are an important component in a wide range of everyday devices.
  Permanent magnets are used, for example, in hard disks, electro motors and direct drive wind turbines.
  Antiferromagnetic compounds find applications in exchange biased films for data storage devices, magnetic random access memories (MRAMs)
and are of interest for spintronics applications~\cite{Sander17,Zelezny14,Baltz18}.
Present high performance magnets like Nd-Fe-B contain a considerable amount of rare-earth elements, which in the past years have been subject to strong price fluctuations.
The development of alternative permanent magnets without the use of critical rare-earth elements is therefore highly important~\cite{Gutfleisch11,Kuzmin14,Skokov18}.
Basic requirements for a good permanent magnet material are a large uniaxial magnetocrystalline anisotropy energy (MAE),
a large saturation magnetization and a high Curie temperature.

High-throughput calculations in the framework of density functional theory (DFT) have become an increasingly important tool for the design of 
new functional materials in the past years~\cite{Curtarolo13,Jain13}. In addition to increased computational resources, this is due to the
development of large databases and new concepts like efficient descriptors. 
A number of studies have focussed on the development of improved materials for batteries~\cite{Hautier11,Saal13},
thermoelectrics~\cite{Madsen06,Wang11,Bera14} or photovoltaics~\cite{Yu12}.

For magnetic materials, despite their technological importance, computational high-throughput studies are scarce.
Bocarsly {\em et al.} performed a screening of potential magnetocaloric materials~\cite{Bocarsly17}, 
and
Els\"asser {\em et al.} performed a screening of potential hard magnetic rare earth lean compounds~\cite{Drebov13,Koerner16,Koerner18}.
Other studies focussed on potential rare earth free permanent magnets~\cite{Edstroem14,Herper18,Vekilova19,Fayyazi19}
 or the prediction of new magnetic compounds~\cite{Sanvito17,Balluff17,Arapan18,Singh18}.
A complete study of the stability, magnetic properties and MAE of potential new magnetic compounds taking into account competing phases and
magnetic structures was to the best of our knowledge not performed up to now.

Theoretical high-throughput descriptions of functional magnetic materials
involve different energy scales like
Coulomb correlations, exchange interactions or magnetocrystalline anisotropy.
Antiferromagnetic (AF) order can have a strong impact on the total energy and phase stability of Mn alloys~\cite{Kuebler88,Balluff17},
which are of interest for hardmagnetic~\cite{Zhang15}, magnetocaloric~\cite{Liu12} as well as
spintronics applications~\cite{Zelezny14}.
The effect of AF interactions on the energetics has been explored in high-throughput calculations
for certain classes of compounds with known magnetic structure prototypes~\cite{Balluff17,Ohmer19,Gao19}
or on selected compounds using evolutionary algorithms~\cite{Arapan18,Payne18} or a symmetry based prioritization~\cite{Horton19}.
Complex (non-collinear) structures and non-ferromagnetic ground states of competing structures in a phase diagram haven't been studied up to now,
as the computational effort is easily enlarged by one order of magnitude, and often suitable magnetic structure prototypes
are not known.
For the calculation of the MAE, which sets an upper limit to the coercivity,
a proper treatment of spin-orbit coupling (SOC) and a high numerical accuracy,
especially a careful convergence of k-space integrals are essential.
Rare earth compounds with localized 4f electrons require a careful treatment
of Coulomb correlations, where accurate methods suitable for a high-throughput screening
still have to be developed.
While individual aspects have been carefully addressed in the literature, the complexity of interactions makes
  a high-throughput prediction of functional magnetic materials still highly demanding.

It has been shown in the literature that interstitials can have a strong impact on the magnetic properties,
including the magneto-crystalline anisotropy energy~\cite{Coey90,Steinbeck96,Zhang16}, the magnetic structure~\cite{Gajdzik00} and magnetostructural
phase transitions~\cite{Lyubina09,Trung10}.
The Cu$_3$Au-structure is one of the most common structure types with more than 200 reported compounds.
Ferromagnetic compounds like Fe$_3$Pt can exhibit a large magnetization, which makes them interesting
for applications in exchange coupled magnets~\cite{Lyubina05,Hai03}.
Due to the cubic symmetry they have however only a relatively small MAE. 
In contrast, antiferromagnetic compounds 
like Mn$_3$Ir can exhibit a huge magneto-crystalline anisotropy~\cite{Szunyogh09},
but have vanishing magnetization.
This is used in exchange biased films, where giant exchange anisotropies are reported~\cite{Imakita04,Kohn13}.
Further, Mn-rich alloys of Cu$_3$Au-prototype have attracted attention due to a large magnetostriction~\cite{Shimizu12}
and Berry curvature related phenomena like a large anomalous Hall effect and a facet-dependent spin Hall effect~\cite{Chen14,Liu18,Zhang16a}.

\begin{figure}
\includegraphics[width=0.5\textwidth]{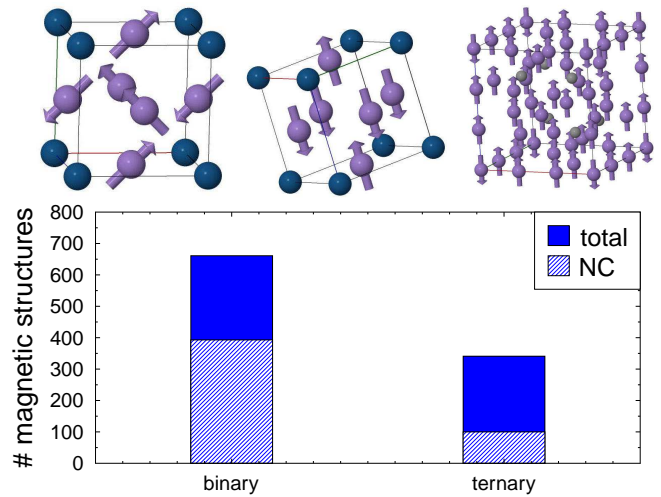}
\caption{\label{FIG:N_MAG}Total number (filled blue) and number of NC (shaded blue) magnetic structures calculated for binary and ternary Mn alloys in this work.
Some examples for automatically generated initial spin structures are shown as well: a non-collinear structure and collinear ferrimagnetic structure for Mn$_3$Ir and
a collinear ferrimagnetic structure for Mn$_{23}$C$_6$ (from left to right).} 
\end{figure}

In this article, we investigate the effect of interstitial N, C and B atoms on the stability and
magnetic properties of 21 experimentally reported
compounds with cubic Cu$_3$Au-structure and the magnetic 3d elements Cr, Mn, Fe, Co, and Ni.
To identify the magnetic ground state of Mn alloys, the energy of different spin structures
with maximal magnetic subgroup symmetry is calculated, which according to 
Ref.~\onlinecite{Gallego16} accounts for the majority of measured magnetic structures 
on the Bilbao Crystallographic Server.
This allows to identify also complex non-collinear (NC) ground states, which are frequently observed in Mn alloys.
In total, about 1000 magnetic configurations are probed to obtain reliable energies for alloys relevant for the phase stability of binary and ternary
Mn alloy systems related to this study~(Fig.\ref{FIG:N_MAG}).
Stable interstitial phases are identified taking into account all possible decomposition reactions.
We show that interstitials can lead to a strong enhancement of the 
magneto-crystalline anisotropy energy in these alloys
and that high-throughput DFT calculations are able to identify hard magnetic materials
even in the presence of complex magnetic interactions.

\section{Computational Details}
Calculations were performed within density functional theory using an
extended version of the high-throughput environment 
HTE~\cite{Opahle12,Opahle13} employing 
the VASP~\cite{Vasp1,Vasp2} and FPLO~\cite{FPLO} codes.
For the present investigation, the HTE has been extended for the calculation of magnetic properties
such as the investigation of different spin configurations and the calculation of the magneto-crystalline anisotropy energy (MAE).
Further, an automatic detection of suitable interstitial positions was implemented.

For structure optimizations and the calculation of formation energies the VASP code~\cite{Vasp1,Vasp2} has been used following a similar approach as in Ref.~\onlinecite{Opahle13}.
For the final results shown here PAW pseudopotentials with a cut-off energy E=500 eV and constant density of k-points corresponding e.g. to
a k-mesh of 14x14x14 k-points in the full Brillouin zone (FBZ) for Mn$_3$Sb were used.
For MAE calculations, the 
full potential local orbital method~\cite{FPLO} (FPLO, version 14-49) was 
used, employing the optimized crystal and (collinear) magnetic structure
obtained in the preceding VASP calculations. In the FPLO  code, accurate 
full potential calculations are combined with the solution
of the 4-component Kohn-Sham-Dirac equation, which implicitly contains 
spin-orbit coupling up to all orders.

As a descriptor for hard magnetic alloys the magnetic force theorem with a moderate k-mesh (up to 20x20x20 k-points in the FBZ)
was used, where starting from a selfconsistent scalar relativistic calculation,
the MAE is evaluated in one-step calculations as the band energy difference 
between different magnetization directions. In addition, selfconsistent
relativistic calculations with up to 32x32x32 k-points in the FBZ were used to check
the MAE for hard magnetic alloys identified in this study.
Further, the magnetic anisotropy of Mn$_3$Ir(N) and Mn$_3$Pt(N) in the non-collinear ground state was calculated using the VASP code
employing the magnetic force theorem. 
Here, a cut-off energy E=400 eV and a k-mesh of 12x12x12  k-points in the FBZ were used.

The results have been cross checked with selfconsistent calculations including spin-orbit coupling using a cut-off energy E=500 eV and k-meshs of 25x25x25 k-points (Mn$_3$IrN) and 13x13x13 k-points (Mn$_3$PtN),
showing a convergence of the anisotropy energy of about 1 meV.

The anomalous Hall conductivity (AHC) is determined by using the WannierTools (WT) code, which is implemented within the framework of tight binding (TB) models~\cite{Wu2017}. 
The maximally localized Wannier functions (MLFWs) TB models are obtained by using the Wannier90 code~\cite{mostofi2008wannier90}. 
We considered the s, p and d orbitals for Mn and Ir atoms, and s plus p orbitals were incorporated for N, 
resulting in 72 and 80 MLFWs for Mn$_3$Ir and Mn$_3$IrN respectively. To evaluate the AHC, the Berry curvature integration is carried out using a uniform kpoints mesh of 401$\times$401$\times$401, which is expressed as following:~\cite{xiao2010berry}
\begin{equation}
\sigma_{xy} = - \frac{e^2}{\hbar} \int \frac{d\mathbf{k}}{\left( 2\pi \right)^3} \sum_{n} f \big[ \epsilon \left( \mathbf{k} \right) -\mu \big] \Omega_{n,xy} \left( \mathbf{k} \right) 
\end{equation} 
\begin{equation}
\Omega_{n,xy}\left( \mathbf{k} \right) = -2\text{Im} \sum_{m \neq n} \frac{{\langle\psi_{\mathbf{k}n}|v_{x}|\psi_{\mathbf{k}m}}\rangle{\langle\psi_{\mathbf{k}m}|v_{y}|\psi_{\mathbf{k}n}}\rangle}{\big[ \epsilon_m\left( \mathbf{k} \right) - \epsilon_n\left( \mathbf{k} \right) \big]^2} \end{equation}
where e is elementary charge, $\mu$ is the chemical potential, $\psi_{n/m}$ denotes the Bloch wave function with energy eigenvalue $\epsilon_{n/m}$,
$v_{x/y}$ is the velocity operator along Cartesian $x/y$ direction, and $f[\epsilon \left( \mathbf{k} \right) -\mu]$ is the Fermi-Dirac distribution function. In this work, the ANC $\alpha_{xy}$ is determined by using the Mott relation, where we considered only the derivative of AHC at the Fermi level d$\sigma_{xy}$/d$\epsilon$, which provides a quantitative measure of the ANC.
\begin{equation}
\alpha_{xy} = - \frac{\pi^2k_B^2T}{3e} \frac {d\sigma_{xy}} {d\epsilon} \big |_{\epsilon=\mu} 
\end{equation}
Unless otherwise mentioned, the generalized gradient approximation (GGA) in the parameterization of Perdew, Burke and
Ernzerhof~\cite{Perdew96} was used.
For calculations in the local spin density approximation (LSDA), which often provides a better description of the spin magnetic moments
for weakly correlated systems,
the parameterization of Perdew and Wang~\cite{Perdew92} is used.

\begin{figure}
\includegraphics[width=0.13\textwidth]{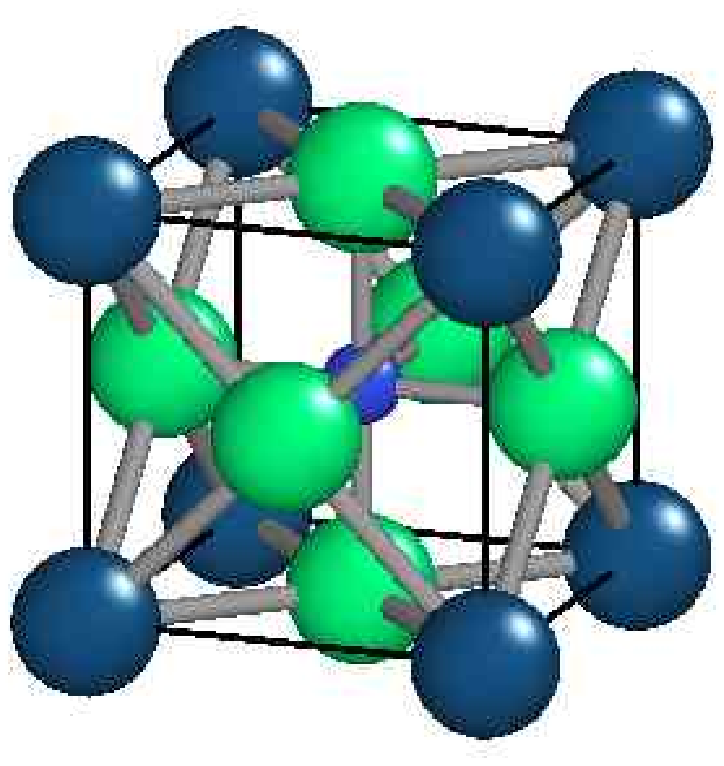}
\includegraphics[width=0.13\textwidth]{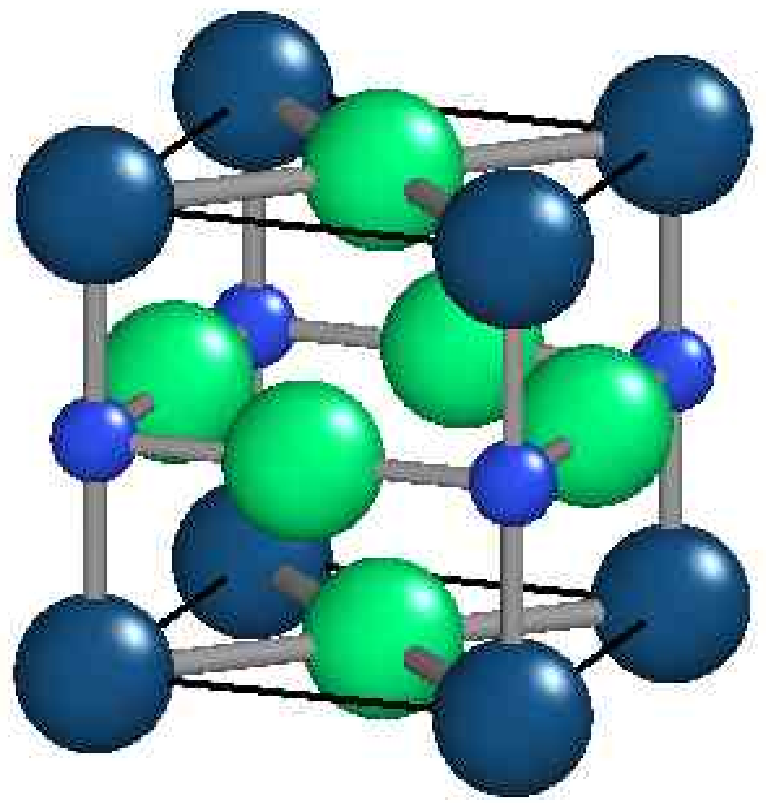}
\includegraphics[width=0.13\textwidth]{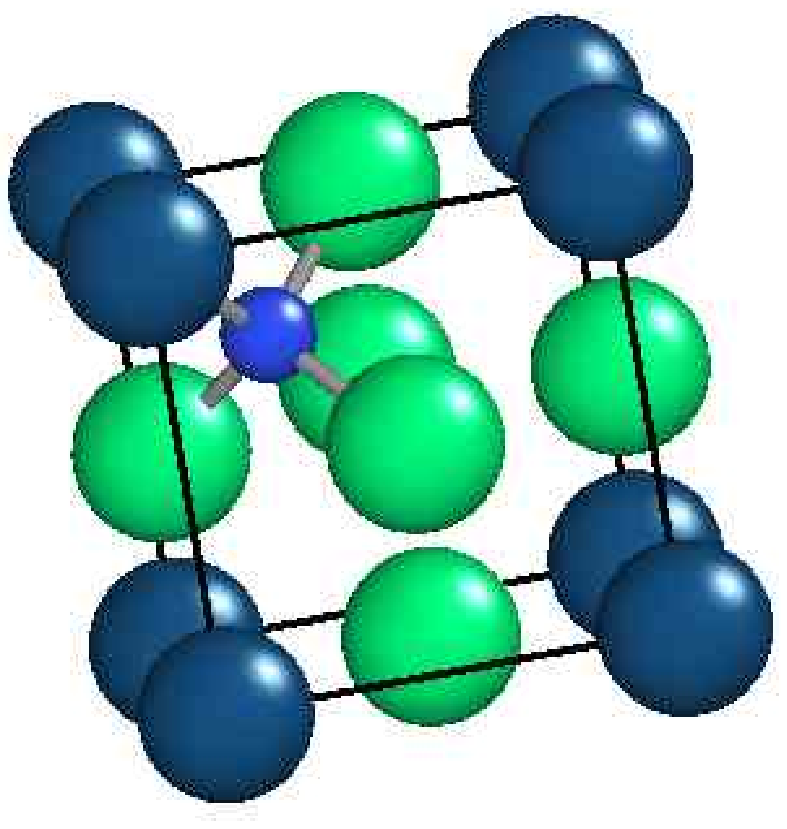}
\caption{\label{FIG:Interstitials}Interstitial positions in the Cu$_3$Au-structure: 1b, 3d and 8g (from left to right).} 
\end{figure}

For the detection of suitable interstitial positions, the unit cell is scanned on a discrete mesh (here 24x24x24 points) for sufficient empty space, taking into account the covalent radii $r_j$ 
of the atoms as tabulated in the Atomic Simulation Environment (ASE)~\cite{ASE,ASE17}.
As condition for suitable positions for the interstitial $I$ we use $d_{\rm min}=\min_{j}\left(d_{Ij}-s(r_I+r_j)\right)>0$ 
for all neighbouring atoms
$j$, where $s=0.7$ is a scaling factor and $d_{Ij}$ is the distance between the interstitial and
the atom $j$.
This approach is similar to the detection of empty spheres used in some electronic structure codes.
Preference can be given to high symmetry positions,
which was done for the present calculations. The resulting interstitial positions for the Cu$_3$Au structure are shown in Fig.~\ref{FIG:Interstitials}.
The interstitial concentration was varied using 2x2x2, 1x2x2, 1x1x2 and 1x1x1
supercells with one interstitial position occupied, 
corresponding to interstitial concentrations of 
3.0, 5.9, 11.1 and 20 at.\%.
An additional slight orthorhombic distortion (2-5\% change of the lattice parameters)  of the interstitial supercells was also tested,
but did not have an effect on the stable interstitial compounds discussed in this manuscript.

The calculation of convex hulls is based on about 2000 alloys (experimental structures and structures close to the convex hull reported in the Materials Project~\cite{MP})
for ternary $M$-$X$-$Z$ systems ($M$=Fe, Co, Ni, Cr and Mn; $Z$=B, C, N; $X$: other element).
For the competing phases, the same numerical setup as for the interstitial compounds 
was used.
For compounds with Mn atoms, where antiferromagnetic couplings are to be expected,
the magnetic ground state was determined for the parent compounds with Cu$_3$Au-structure and their stable interstitial alloys.
To ensure reliable phase diagrams for the Mn systems, the magnetic ground state was also determined for the
experimentally reported competing phases and some of the energetically low lying phases reported in the Materials Project.
Depending on the size and symmetry of the cell, we used spin configurations compatible with the
maximal magnetic subgroups with propagation vector {\bf q}=(0,0,0), (1/2,1/2,0), (0,0,1/2) or
alternatively an antiferromagnetic coupling of sublattices to probe the magnetic ground state.

In case of Mn$_3$Ir with Cu3Au-prototype this results (after removal of symmetry equivalent configurations) in three collinear (FM, AF and ferrimagnetic) and one non-collinear
 initial spin structures for {\bf q}=(0,0,0), which is sufficient
  to obtain the experimentally reported ground state.
  It is noteworthy that the experimentally reported magnetic structure of Mn$_3$Ir does not have maximal magnetic subgroup symmetry,
  but is compatible with the symmetry of the second generation of maximal subgroups (maximal subgroup $R3m'$ (\#166.101) of the maximal subgroup $Pm\bar{3}m'$ (\#221.95)). 
  Nevertheless, the NC initial structures relax to the correct ground state, which underpins the high efficiency of the approach.
  We explicitely checked that the second generation of maximal subgroups 
  results in
  the same magnetic ground state.
  The AF coupling of sublattices corresponds to the highest ranked magnetic structures in the prioritization scheme proposed in
  Ref.~\onlinecite{Horton19} and is especially useful when the paramagnetic structure has a relatively large number of
  inequivalent magnetic atoms $N_{\rm AF}$ which are allowed to couple AF.
  The number of possible magnetic configurations at this prioritization level is $2^{N_{\rm AF}-1}$, which results for instance for
  Mn$_7$C$_3$ of hexagonal Cr7C3,hP80,186-prototype 
  (with 56 Mn atoms in the unit cell and $N_{\rm AF}=8$) in 128 possible magnetic configurations.
While this approach does not ensure that the correct magnetic ground state is found in all cases,
we believe it should give a reliable basis for the prediction of the phase stability in these
systems.
For the remaining alloys a ferromagnetic magnetic configuration was assumed.

\section{Results of High-Throughput Screening}
\subsection{Parent compounds in Cu$_3$Au-structure}

\begin{table}
\begin{tabular}{c|c|c|c|c|c}
  alloy & $a_{\rm exp}$[\AA] & $a_{\rm GGA}$[\AA] & $E_f$[eV/at.] & $\Delta E_{h}$[eV/at.]  & $M$[$\mu_B$/f.u.]\\ \hline
Co$_{3}$Al  &  3.66  &  3.57  & -0.175  & 0.124  & 3.86  \\   
Co$_{3}$Ta  &  3.65  &  3.63  & -0.241  & 0.001  & 0.02  \\   
Co$_{3}$Ti  &  3.62  &  3.60  & -0.257  & 0.000  & 2.71  \\   
Fe$_{3}$Ga  &  3.70  &  3.66  & -0.133  & 0.000  & 6.98  \\   
Fe$_{3}$Ge  &  3.68  &  3.63  & -0.099  & 0.010  & 6.37  \\   
Fe$_{3}$Pt  &  3.77  &  3.73  & -0.076  & 0.041  & 8.46  \\   
Fe$_{3}$Sn  &  3.87  &  3.82  & 0.086  & 0.088  & 6.94  \\   
Mn$_{3}$Ir  &  3.77  &  3.71  & -0.215  & 0.000  & 0.01$^{*}$  \\   
Mn$_{3}$Pt  &  3.84  &  3.74  & -0.251  & 0.000  & 0.00$^{*}$  \\   
Mn$_{3}$Rh  &  3.81  &  3.71  & -0.142  & 0.000  & 0.00$^{*}$  \\   
Mn$_{3}$Sb  &  4.00  &  3.91  & 0.076  & 0.086  & 0.04$^{*}$  \\   
Ni$_{3}$Fe  &  3.55  &  3.55  & -0.088  & 0.000  & 4.72  \\   
Ni$_{3}$Ga  &  3.58  &  3.59  & -0.289  & 0.000  & 0.79  \\   
Ni$_{3}$Ge  &  3.57  &  3.57  & -0.292  & 0.000  & 0.05  \\   
Ni$_{3}$In  &  3.75  &  3.75  & -0.052  & 0.045  & 0.69  \\   
Ni$_{3}$Mn  &  3.59  &  3.56  & -0.102  & 0.000  & 4.85  \\   
Ni$_{3}$Si  &  3.51  &  3.51  & -0.462  & 0.000  & 0.04  \\   
Ni$_{3}$Sn  &  3.74  &  3.74  & -0.188  & 0.000  & 0.06  \\   
Pt$_{3}$Co  &  3.83  &  3.88  & -0.061  & 0.000  & 2.91  \\   
Pt$_{3}$Cr  &  3.88  &  3.91  & -0.254  & 0.000  & 2.60  \\   
Rh$_{3}$Cr  &  3.79  &  3.78  & -0.122  & 0.000  & 1.24  \\   
\end{tabular}
\caption{\label{TAB:Cu3Au}Properties of parent compounds in Cu$_3$Au-structure.
Experimental ($a_{\rm exp}$) and calculated lattice parameters ($a_{\rm GGA}$), the calculated formation
energy $E_f$ and distance from the convex hull $\Delta E_{h}$ and the total magnetic
moment $M$ are shown. Non-collinear magnetic structures are marked with $^{*}$.
} 
\end{table}
Tab.~\ref{TAB:Cu3Au} shows the calculated properties of the 21 parent compounds with Cu$_3$Au structure.
The calculated lattice parameters are in good agreement with the reported experimental values,
with the largest deviation of about 2.6\% for Mn$_3$Rh and Mn$_3$Pt.
With exception of Mn$_3$Sb and Fe$_3$Sn the calculated formation energy is negative for all compounds,
as expected from formation of alloys in nature.
Mn$_3$Sb has only been recently synthesized by high pressure synthesis~\cite{Yamashita03}
and the Cu$_3$Au-structure of Fe$_3$Sn was reported in high pressure studies,
whereas the ambient pressure phase is hexagonal, in agreement with our calculations  (see Supplementary). 
For 18 of the 21 compounds the calculated distance from the convex hull $\Delta E_{h}$ is below 50 meV/atom, which has been proposed in 
Ref.~\onlinecite{Opahle13} as a simple descriptor for potentially stable or meta stable alloys in high-throughput calculations. 
In addition to the above mentioned Mn$_3$Sb and Fe$_3$Sn alloys, only Co$_3$Al is above this criterion, which shows the largest $\Delta E_{h}= 124$ meV/atom among the calculated compounds.
A Cu$_3$Au-structure of Co$_3$Al has been discussed in the literature, but could not be stabilized experimentally~\cite{Ellner92}. 
This result is confirmed by our study, where only the Al rich phases AlCo, 
Al$_{5}$Co$_2$, 
Al$_{13}$Co$_4$, 
and Al$_{9}$Co$_2$ 
are found to be stable.
A complete list of the alloys relevant for the calculation of the convex hull in Tab.~\ref{TAB:Cu3Au} is given in the supplementary material.

\begin{figure}
\includegraphics[width=0.45\textwidth]{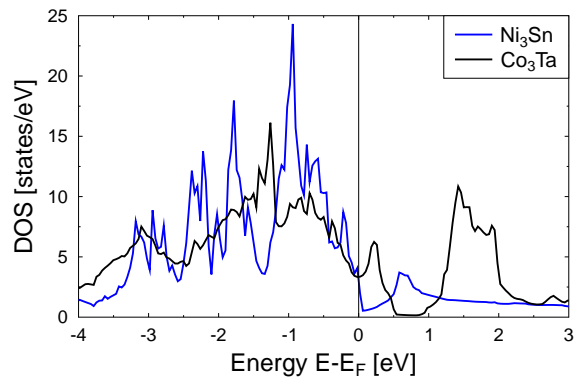}
\caption{\label{FIG:DOS-Cu3Au}Total DOS of Co$_3$Ta and  Ni$_3$Sn for the experimental lattice parameters calculated with FPLO using GGA.} 
\end{figure}

For Ni$_3X$ ($X$=Si, Ge, Sn) and Co$_3$Ta the Fermi energy lies in a pseudo gap or valley of the density of states (DOS), see Fig.~\ref{FIG:DOS-Cu3Au}, resulting in a non-magnetic or only weakly ferromagnetic ground state 
despite the presence of the magnetic elements Ni and Co.
For Mn$_3X$ ($X$=Rh, Ir, Pt and Sb) the calculated ground state is non-collinear. The antiferromagnetic coupling of the neighbouring Mn atoms can not be fully satisfied in the Cu$_3$Au structure, resulting in a frustrated triangular spin structure with compensated magnetic moments,
see for instance  Ref.~\onlinecite{Szunyogh09} for a more detailed discussion.
For $X$=Rh, Ir, Pt  this result is in agreement with experiment~\cite{Kren68,Tomeno99} and earlier calculations~\cite{Szunyogh09}.
In case of Mn$_3$Pt a transition from a triangular to a collinear AF magnetic structure with a doubled unit cell along $c$ was observed around 365 K in neutron scattering experiments~\cite{Kren68}, while a similar transition was
not observed for Mn$_3$Rh and Mn$_3$Ir. For Mn$_3$Sb, the energies of NC, ferrimagnetic and ferromagnetic (FM) magnetic structures are within less than 35 meV/atom. Experimentally, a collinear magnetic structure with almost compensating Mn magnetic moments
was deduced from neutron scattering experiments~\cite{Ryzhkovskii11},
while more recent M\"ossbauer studies suggest a triangular magnetic structure~\cite{Budzynski14}.

\begin{figure}
\includegraphics[width=0.45\textwidth]{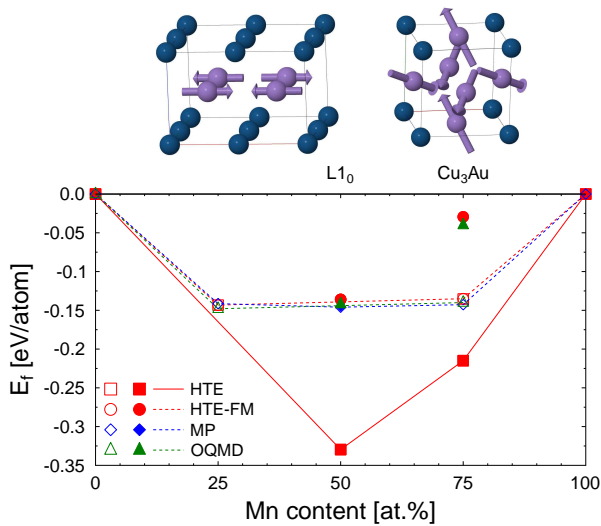}
\caption{\label{FIG:PD-Mn-Ir}
  Calculated formation energies (red squares) and convex hull (solid red line) for the magnetic ground state of Mn-Ir alloys.
  For comparison the convex hull (dashed lines) and formation energies obtained assuming a ferromagnetic structure
  from our calculations (red circles), the Materials Project (blue diamonds) and the Open Quantum Materials Database
  (green triangles) are shown. Formation energies for the experimentally reported L1$_0$- and Cu$_3$Au-crystal structures
  obtained with the different calculation methods are plotted with filled symbols.
  Top: calculated magnetic groundstate structures for MnIr and Mn$_3$Ir. Note that SOC was not included here
  and that the spin-quantization axis has been rotated to match the results of calculations with SOC.
} 
\end{figure}

The energy difference $\Delta E_{\rm FM}$ between the magnetic ground state and a ferromagnetic state
can be as large as 243 meV/atom for Mn$_3$Pt.
For some of the competing phases the calculated $\Delta E_{\rm FM}$ is of similar magnitude, while  $\Delta E_{\rm FM}$ is
trivially zero for compounds with a ferromagnetic ground state.
The magnetic structure can thus have a deep impact on the calculated phase stability, going beyond typical effects due to
finite temperature.
This is exemplarily shown in Fig.~\ref{FIG:PD-Mn-Ir} for binary Mn-Ir alloys.
Assuming a ferromagnetic spin alignment a hexagonal MnIr$_3$ phase, L1$_0$-MnIr and a hexagonal Mn$_3$Ir phase
are on or nearly on the convex hull, in good agreement with results obtained from the Materials Project~\cite{MP}
and the Open Quantum Materials Database OQMD~\cite{OQMD}.
In contrast, calculations with the magnetic ground state yield the experimentally observed  L1$_0$-MnIr and
Cu$_3$Au-structure of Mn$_3$Ir as stable phases on the convex hull.
For L1$_0$-MnIr and Cu$_3$Au-Mn$_3$Ir the magnetic ground state search resulted in a collinear AF and a NC triangular AF structure, respectively, in agreement with experiment~\cite{Tomeno99,Selte68}. This is accompanied by a significant lowering of the formation energy
of nearly 0.2 meV/atom. 
In contrast, the magnetic ground state search for the two (experimentally not observed) hexagonal phases
did not have an effect on the calculated formation energy.
A similar impact of the magnetic structure on the calculated phase stability as for Mn-Ir is found also in other Mn alloy systems
and will be discussed in more detail elsewhere.

\subsection{Effect of interstitials on structural and magnetic properties}

\begin{figure}
  \includegraphics[width=0.45\textwidth]{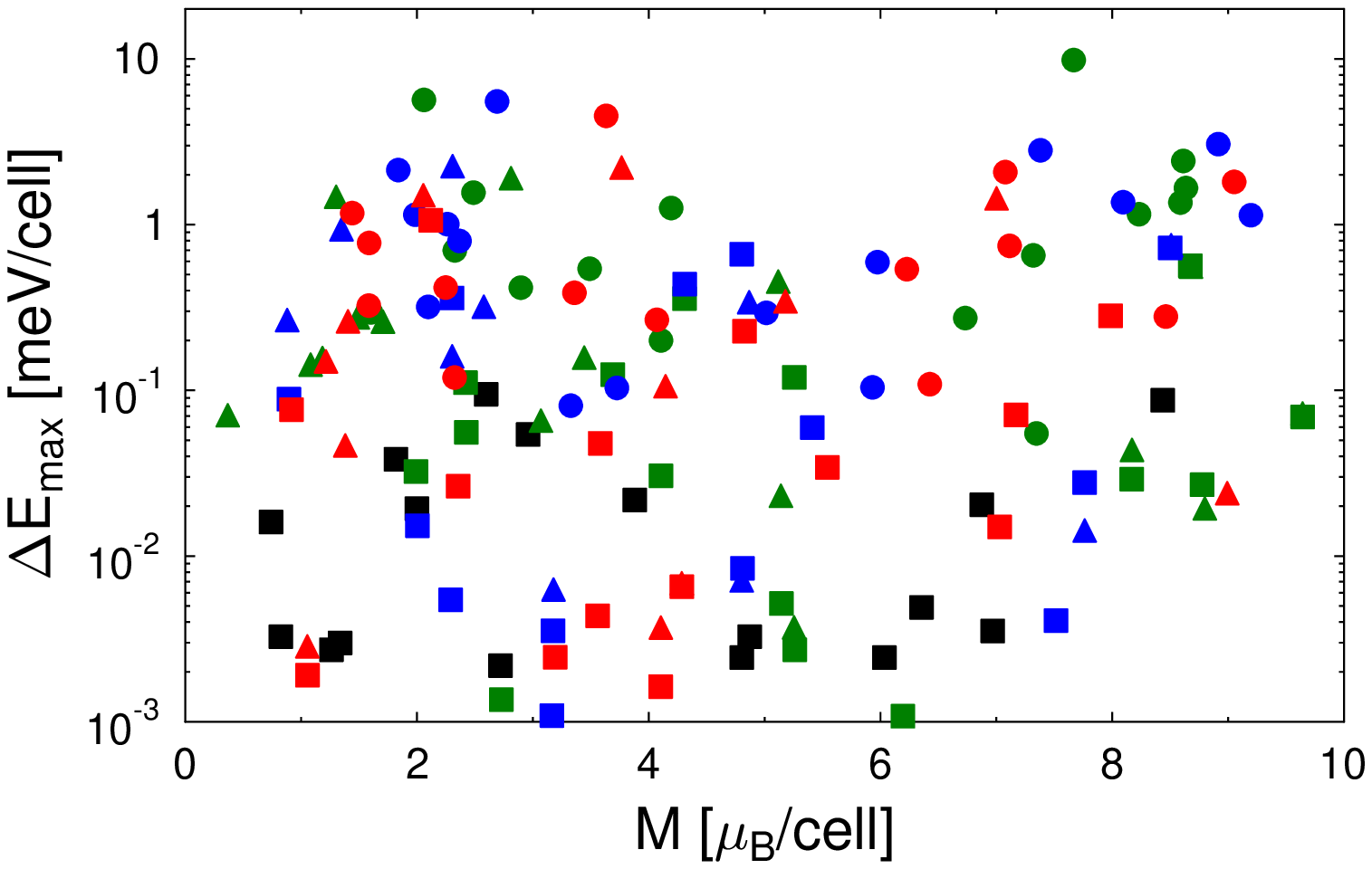}
  \includegraphics[width=0.45\textwidth]{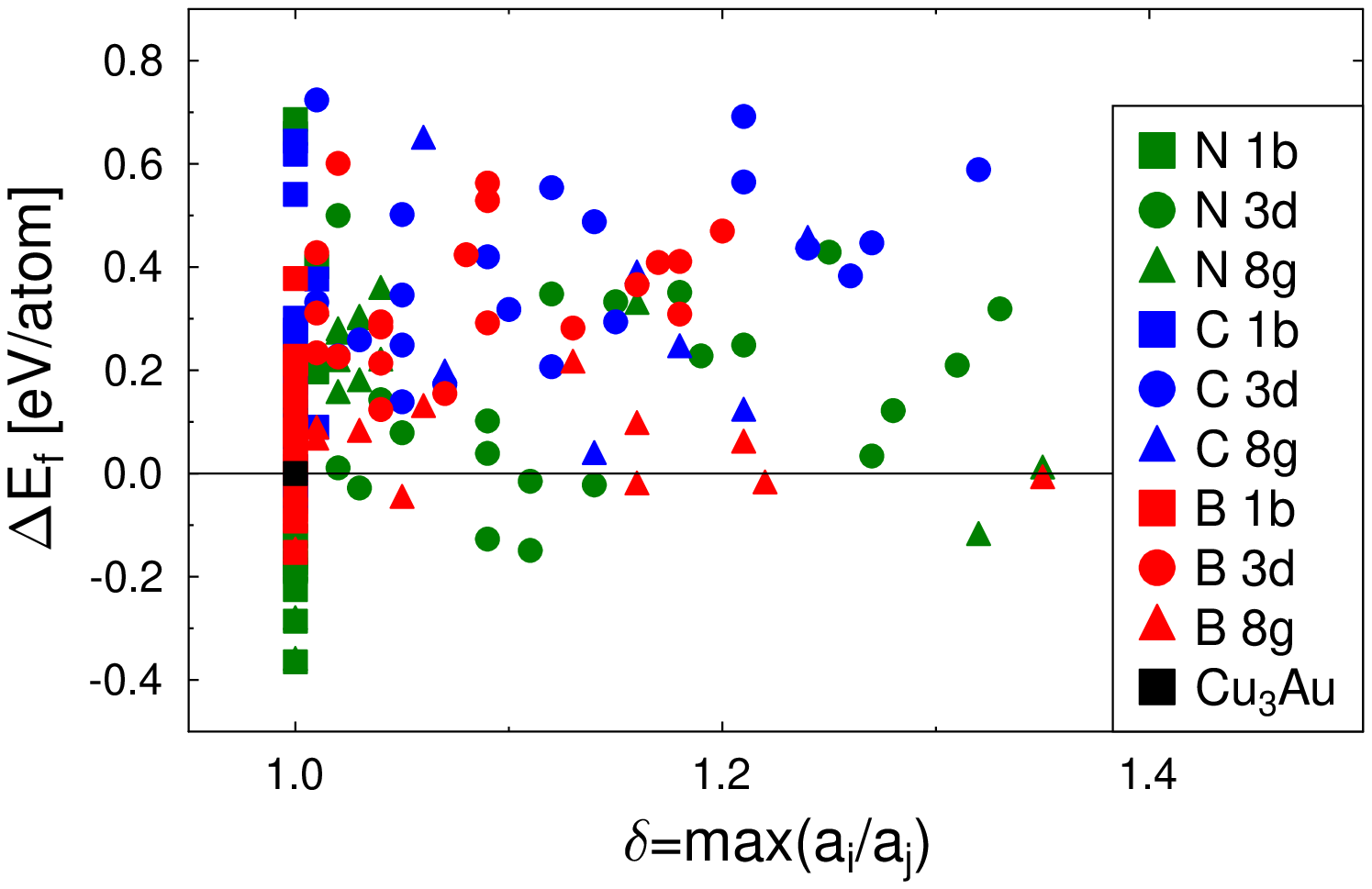}
  \caption{\label{FIG:MAEscatter}Maximum energy difference $\Delta E_{\rm max}=\max(E_{i}-E_{j})$ between different magnetization directions
    vs. total magnetic moment $M$ (top) and 
    maximal change of the ratio of lattice parameters $\delta =\max(a_i/a_j)$ vs. change in the formation energy $\Delta E_f$ (bottom).
    Data are shown for parent compounds in the Cu$_3$Au-structure (filled black squares) and interstitial structures with B (red), C (blue)
    and N (green). The interstitial 1b, 3d and 8g positions are marked with filled squares, circles and triangles, respectively.} 
\end{figure}

Fig.~\ref{FIG:MAEscatter} shows the calculated MAE (maximum energy difference between magnetization directions
[001], [010], [100], [110], and [111]) for the Cu$_3$Au parent compounds and interstitial alloys with one N, C or B atom per f.u., assuming a ferromagnetic alignment of the spins.
It can be seen that interstitials can induce huge MAEs up to about 2 meV/atom, one to two orders higher in magnitude than in their cubic parent compounds. 
N, C and B are similarly efficient in inducing these large MAEs. 
The largest MAEs are calculated for the 3d interstitial position, which leads to a tetragonal distortion of the lattice
and therefore is promising to achieve uniaxial MAEs.
There can be also a significant MAE between the [001], [110] and [111] directions of up to about 0.2 meV/atom for the cubic compounds with heavy 5d elements
(while the energy difference between [001], [010] and [100] is zero due to symmetry).
It should be noted however, that neither stability nor potential antiferromagnetic spin configurations are considered in this plot.
The MAEs shown in Fig.~\ref{FIG:MAEscatter} thus rather serve to illustrate the potential of interstitials in inducing large MAEs,
while for the majority of the data points an experimental realization will not be possible.

This becomes also clear from the lower plot of Fig.~\ref{FIG:MAEscatter}, which shows the formation energy difference $\Delta E_f$ with respect
to the Cu$_3$Au parent compound in relation to the induced structural distortion for different interstitial atoms and positions.
For the majority of data points $\Delta E_f$ is positive indicating that the interstitial alloy is unlikely to form, at least at the interstitial concentration of 20 at.\%
considered in the plot. Further, in most cases the 1b interstitial position, which does not break the cubic symmetry, is most stable. 
In most cases this means that a significant enhancement of the MAE is not expected. Noteworthy exceptions, where the spin structure breaks the cubic symmetry, 
will be discussed further below.
Significant structural distortions are found for the 3d and 8g interstitial positions with negative $\Delta E_f$ in some cases, suggesting
that the interstitial alloys are likely to form. 
For the 8g interstitial position, which is least favourable from geometrical considerations, we often find relaxation to a structure corresponding
to the 1b or 3d position, which provide a higher volume for the interstitial atom.

\begin{table}
\begin{tabular}{c|c|c|c|c|c}
parent & interst. & $c_{\rm max}$ & $E_f$ [eV/at.] & $\Delta E_{h}$[eV/at.]  & $M$[$\mu_B$/at.] \\ \hline
Co$_{3}$Al  & C 1b  & 20.0  & -0.230  & 0.010  & 0.00  \\   
Co$_{3}$Ta  & B 1b  & 11.1  & -0.272  & 0.044  & 0.00  \\   
Co$_{3}$Ta  & C 1b  & 11.1  & -0.211  & 0.035  & 0.00  \\   
Co$_{3}$Ti  & B 1b  & 11.1  & -0.324  & 0.043  & 0.31  \\   
Co$_{3}$Ti  & C 1b  & 11.1  & -0.302  & 0.003  & 0.22  \\   
Fe$_{3}$Ga  & B 1b  & 5.9  & -0.124  & 0.047  & 1.50  \\   
Fe$_{3}$Ga  & C 1b  & 20.0  & -0.063  & 0.044  & 0.63  \\   
Fe$_{3}$Ga  & N 1b  & 20.0  & -0.255  & 0.016  & 1.05  \\   
Fe$_{3}$Ge  & N 1b  & 5.9  & -0.096  & 0.040  & 1.32$^{\dagger}$  \\   
Fe$_{3}$Pt  & N 1b  & 20.0  & -0.263  & 0.000  & 1.63  \\   
Fe$_{3}$Sn  & N 1b  & 20.0  & -0.104  & 0.037  & 1.02  \\   
Mn$_{3}$Ir  & N 1b  & 20.0  & -0.359  & 0.026  & 0.00$^{*}$  \\   
Mn$_{3}$Pt  & N 1b  & 20.0  & -0.471  & 0.000  & 0.00$^{*}$  \\   
Mn$_{3}$Rh  & N 1b  & 20.0  & -0.356  & 0.000  & 0.00$^{*}$  \\   
Mn$_{3}$Sb  & N 1b  & 20.0  & -0.226  & 0.036  & 0.55$^{\dagger}$  \\   
Ni$_{3}$Fe  & N 3d  & 5.9  & -0.063  & 0.038  & 1.05  \\   
Ni$_{3}$Ga  & B 1b  & 5.9  & -0.305  & 0.014  & 0.00  \\   
Ni$_{3}$Ga  & C 1b  & 5.9  & -0.273  & 0.006  & 0.00  \\   
Ni$_{3}$Ga  & N 1b  & 11.1  & -0.258  & 0.036  & 0.00  \\   
Ni$_{3}$In  & B 1b  & 20.0  & -0.205  & 0.049  & 0.00  \\   
Ni$_{3}$In  & C 1b  & 20.0  & -0.040  & 0.038  & 0.00  \\   
Ni$_{3}$In  & N 1b  & 20.0  & -0.173  & 0.000  & 0.00  \\   
Ni$_{3}$Mn  & B 1b  & 5.9  & -0.110  & 0.041  & 1.09  \\   
Ni$_{3}$Mn  & C 1b  & 5.9  & -0.071  & 0.025  & 1.08  \\   
Ni$_{3}$Mn  & N 3d  & 11.1  & -0.113  & 0.044  & 0.95  \\   
Ni$_{3}$Sn  & B 1b  & 5.9  & -0.196  & 0.038  & 0.00  \\   
Ni$_{3}$Sn  & C 1b  & 5.9  & -0.134  & 0.043  & 0.00  \\   
Ni$_{3}$Sn  & N 1b  & 11.1  & -0.135  & 0.045  & 0.00  \\   
Pt$_{3}$Co  & B 8g  & 11.1  & -0.069  & 0.048  & 0.59  \\   
\end{tabular}
\caption{\label{TAB:StableInterstitials}Interstitial alloys with a distance from the convex hull  $\Delta E_{h}< 50$ meV/atom.
The parent compound with Cu$_3$Au-structure, the interstitial atom and position and the highest interstitial concentration $c_{\rm max}$ 
with $\Delta E_{h}< 50$ meV/atom (with three different values of the interstitial concentrations: 5.9, 11.1 and 20) 
in at.\% 
are shown along with the calculated
formation energy $E_f$, $\Delta E_{h}$ and the total magnetic moment $M$ per atom. Non-collinear magnetic structures and antiferromagnetic/
ferrimagnetic structures are indicated with $^{*}$ and $^{\dagger}$, respectively.
}
\end{table}
\subsection{Stable interstitial phases}
Tab.~\ref{TAB:StableInterstitials} lists interstitial alloys, where the distance to the convex hull $\Delta E_{h}$ is below 50 meV/atom
for an interstitial concentration of more than 5 at.\%, suggesting that a substantial amount of interstitial atoms can be incorporated in the
parent alloy with Cu3Au-structure. This criterion is matched for 29 parent/interstitial combinations, including known antiperovskites like 
Mn$_3X$N ($X$=Rh, Ir, Pt and Sb)~\cite{Fruchart78,Shimizu12}, 
Ni$_3$InN~\cite{Cao09}, Fe$_3$PtN~\cite{Wiener55}, Co$_3$AlC~\cite{Huetter58}, 
Fe$_3$GaN~\cite{Houben09} and Fe$_3$SnN~\cite{Scholz15}.
Only the most stable interstitial position is listed.
In some cases more than one interstitial position lowers the formation energy, 
suggesting a partial occupation of the interstitial sites. 
This  is for instance the case for Ni$_{3}$Mn, where the N 3d and 1b positions are close in energy
and both of them fulfill the $\Delta E_{h}$ criterion for 5.9 at.\% N 
concentration.

The largest gain in formation energy with respect to the parent compound $\Delta E_f\approx -0.3$ eV/atom is found for Mn$_3$SbN, 
where the 'interstitial' alloy~\cite{Fruchart78} was reported more than 20 years 
before the meta stable 'parent' alloy Mn$_3$Sb~\cite{Yamashita03}.
For the majority of interstitial alloys, stability is accompanied by a negative $\Delta E_f$. 
However, even a large  negative $\Delta E_f$ does not ensure formation of a stable alloy. This is the case for
instance for Co$_3$AlN and Co$_3$TiN,
where the distance from the convex hull is 0.3-0.4 eV/atom, despite $\Delta E_f$  values of about -0.16 and -0.13 eV/atom, respectively.
Vice versa, a positive $\Delta E_f$ does not prevent the formation of a stable interstitial alloy, as in the case of Fe$_3$GaC$_x$.

For the majority of interstitials, the 1b position is energetically preferred. 
The 3d position, which is especially interesting for hard magnetic applications as it breaks the cubic symmetry and leads to a tetragonal distortion of the lattice, 
is energetically most preferable for Ni$_3$FeN$_x$, Ni$_3$MnN$_x$, Co$_3$TaN$_x$ 
and Co$_3$TiN$_x$.
In all cases substantial MAEs between -0.67 and 1.3 meV/cell are induced for
$x=1$ (20 at.\% N). The largest absolute value of 1.3 meV/cell is found for
Ni$_3$MnN, albeit with an easy plane magnetization.
However, for Co$_3$TaN$_x$ and Co$_3$TiN$_x$ only a minor amount 
of interstitials is expected to be incorporated due to the formation of extraordinary stable Ta-N and Ti-N phases (see Supplementary). 
For realistic N concentrations (11 at.\% N, $x= 0.5$),
Ni$_3$MnN$_x$ shows a significant uniaxial MAE
around -100 $\mu$eV/atom, comparable to hcp-Co.
For Ni$_3$FeN$_x$, the MAE is small ($\approx$ 15 $\mu$eV/atom) and depends on the choice of supercell. 

Tab.~\ref{TAB:MAE} lists stable alloys with large uniaxial MAE for the most stable collinear
magnetic structure obtained in the preceding VASP calculations. For comparison, also results obtained within LSDA (employing the same
GGA optimized structure) are shown. For both GGA and LSDA, the listed alloys show a large uniaxial MAE, but the absolute values
can differ significantly.
While GGA typically yields lattice parameters in better agreement with experiment, the local spin
magnetic moments are often better described within LSDA. 
As the MAE depends on subtle details of the electronic structure,
this can have a strong impact on the calculated MAE~\cite{Ener19}, as is the case here for Mn$_3$Ir, Mn$_3$PtN and Mn$_3$SbN.
According to the table, Mn$_3X$N and Mn$_3X$ ($X$=Rh, Ir, Pt and Sb) would have a huge potential for application as hard magnets, provided they could be stabilized in a ferrimagnetic magnetic structure.
The largest MAE of about $-1.3$ meV/atom is obtained for Mn$_3$IrN. Comparison with Tabs.~\ref{TAB:Cu3Au} and ~\ref{TAB:StableInterstitials} shows however that with exception of Mn$_3$SbN these large MAEs can not be readily used in
bulk magnets, as the ground state is non-collinear with zero net magnetization.
However, the strong MAE could be used in exchange coupled multilayer structures, where a hard magnetic antiferromagnetic layer
is coupled to a soft magnetic layer with large magnetization and high Curie temperature.
Moreover, as Mn$_3$SbN shows, a ferrimagnetic ground state may be induced either by variation of the interstitial concentration
or epitaxial strain in a thin film. This will be further explored in Sec.~\ref{SEC:Discussion}.

Comparison of Tab.~\ref{TAB:StableInterstitials} and Tab.~\ref{TAB:Cu3Au} shows some remarkable effects of the interstitials on the magnetic properties.
While the ground states of Co$_3$Al is ferromagnetic, Co$_3$AlC is found to be non-magnetic.
Similar, the total magnetic moment of Co$_3$Ti is strongly reduced by interstitial B,
resulting in a non-magnetic ground state for (unstable) Co$_3$TiB. 
On the contrary,
Co$_3$Ta is nonmagnetic, while Co$_3$TaN would be magnetic if it could be synthesized. 
This shows that interstitials can be very efficient to tune magnetic
transitions, which makes interstitial alloying interesting also for magnetocaloric applications.

\begin{table}
\begin{tabular}{l|l|c|c}
  alloy & interst. & $M_{\rm GGA}$/$M_{\rm LSDA}$ & MAE$_{\rm GGA}$/MAE$_{\rm LSDA}$ \\
  && [$\mu_B$/atom] & [meV/atom]\\\hline 
Mn$_{3}$Ir  & -- --  & 0.37$^{\dagger}$/0.30$^{\dagger}$&-0.19/-0.43\\ 
Mn$_{3}$IrN  & N 1b  & 0.54$^{\dagger}$/0.49$^{\dagger}$&-1.32/-1.27\\ 
Mn$_{3}$Pt  & -- --  & 0.36$^{\dagger}$/0.32$^{\dagger}$&-0.09/-0.11\\ 
Mn$_{3}$PtN  & N 1b  & 0.63$^{\dagger}$/0.57$^{\dagger}$&-0.16/-0.22\\ 
Mn$_{3}$Rh  & -- --  & 0.43$^{\dagger}$/0.37$^{\dagger}$&-0.10/-0.12\\ 
Mn$_{3}$RhN  & N 1b  & 0.56$^{\dagger}$/0.51$^{\dagger}$&-0.31/-0.29\\ 
Mn$_{3}$SbN  & N 1b  & 0.55$^{\dagger}$/0.52$^{\dagger}$&-0.20/-0.13\\ 
Ni$_{3}$MnN$_{0.5}$  & N 3d  & 0.96/0.90&-0.09/-0.10  
\end{tabular}
\caption{\label{TAB:MAE}Stable alloys with large uniaxial MAE=$E_{[001]}-E_{[100]}$.
  The total magnetic spin moment $M$ and the MAE calculated within GGA/LSDA
  are listed.
  A collinear spin arrangement is assumed.
  Compounds with a ferrimagnetic structure are indicated with $^{\dagger}$ on the magnetic moment.
}
\end{table}
\section{Discussion\label{SEC:Discussion}}
In the previous section we have shown that interstitial atoms can have a
strong impact on the stability, magnetic properties and MAE of compounds.
While in principle huge MAEs are possible (see Fig.~\ref{FIG:MAEscatter}),
{\em stable} compounds with the necessary properties for a good permanent
magnet are rare, in line with experimental experience.
Among the 63 parent/interstitial combinations investigated in this study,
Ni$_{3}$MnN$_{x}$ with $x\approx 0.5$ was identified as a stable
ferromagnetic alloy with magnetic properties comparable to hcp-Co
at reduced cost.
Further, Mn$_{3}$SbN was identified as a ferrimagnet with large MAE,
in agreement with a large magnetostriction observed experimentally in this
compound~\cite{Shimizu12}.
It can be expected that an extended screening of more parent/interstitial
combinations leads to even more promising candidates for hard magnetic
applications.

\begin{figure}
\includegraphics[width=0.45\textwidth]{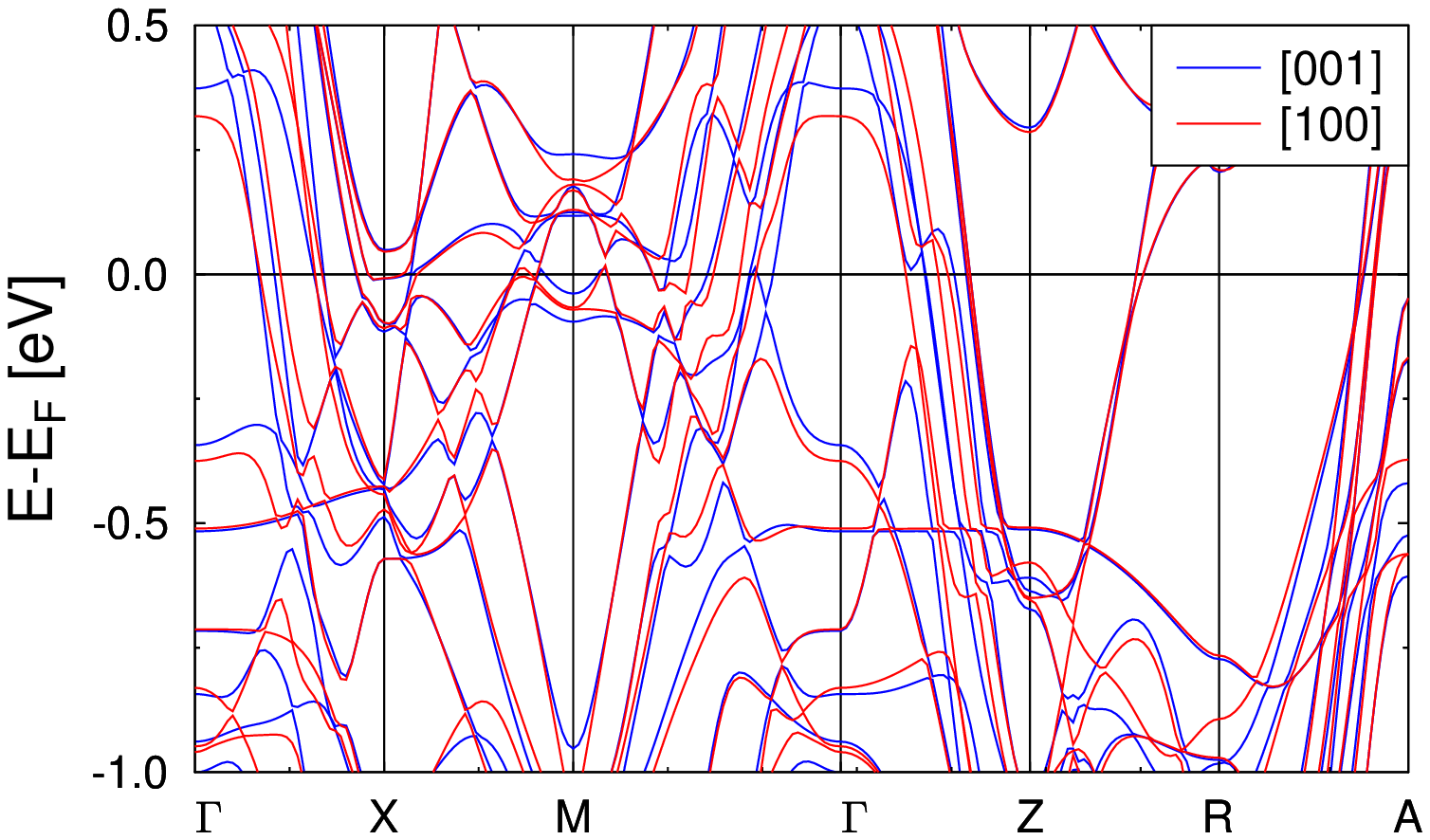}
\includegraphics[width=0.45\textwidth]{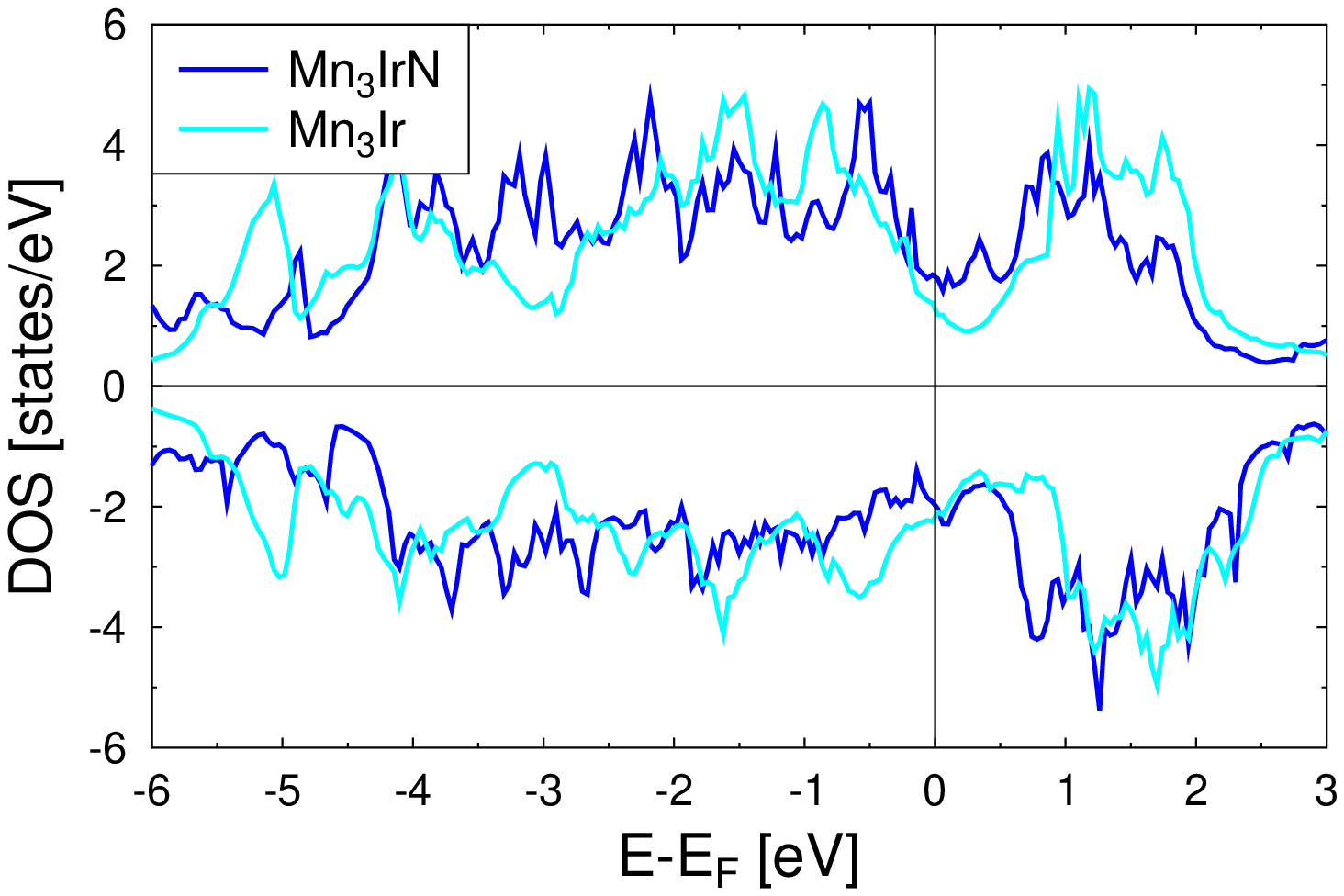}
\caption{\label{FIG:DOSBSMn3IrN}Band structure (top) of ferrimagnetic Mn$_3$IrN for the easy [001] axis (blue) and
  the hard [100] axis (red).
Comparison of the DOS of ferrimagnetic Mn$_3$IrN (blue) and ferrimagnetic Mn$_3$Ir (cyan) for the easy [001] axis (bottom).
For simplicity $c/a$ in both graphs was set to 1.
}
\end{figure}

The most striking observation in this study is however the huge MAE
in Mn$_3X$N$_x$ ($X$=Rh, Ir, Pt, and Sb) and the strong impact of
interstitial N on the MAE in these compounds, especially for Mn$_3$IrN.
Fig.~\ref{FIG:DOSBSMn3IrN} shows the calculated DOS of Mn$_3$IrN 
and Mn$_3$Ir. Interstitial N induces changes in the electronic structure
in the range of a few hundred meV, larger than the changes
in the electronic structure due to a change of the magnetization direction
from the  easy [001] axis to the hard [100] axis due to spin-orbit coupling,
also shown for comparison in Fig.~\ref{FIG:DOSBSMn3IrN}.
The changes in the electronic structure lead also to a change of the
orbital moment anisotropy of Ir ($\Delta\mu_l=0.046 \mu_B$ for Mn$_3$Ir;
$\Delta\mu_l=0.073 \mu_B$ for Mn$_3$IrN) and Mn 
($\Delta\mu_l=0.052 \mu_B$ for Mn$_3$Ir; $\Delta\mu_l=-0.085 \mu_B$ 
for Mn$_3$IrN), which is closely related to the MAE.
Thus, interstititial N has an impact on the electronic structure and MAE
going far beyond structural distortions, in agreement with earlier 
investigations~\cite{Zhang16}.

\begin{figure}
\includegraphics[width=0.45\textwidth]{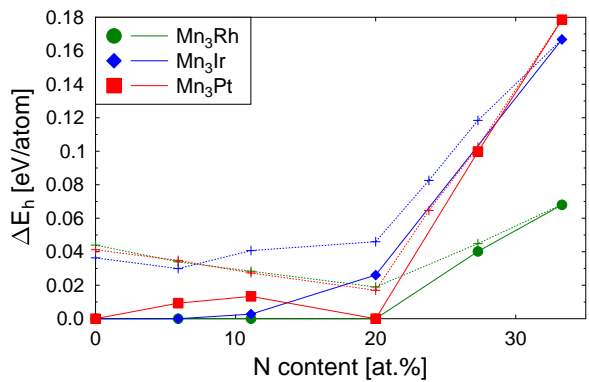}\\
\caption{\label{FIG:Mn3XvsN}
Distance from convex hull $\Delta E_h$ 
vs. N concentration for Mn$_3X$N$_x$ ($X$=Rh, Ir, Pt)
(filled symbols: NC, plus: ferrimagnetic). 
}
\end{figure}

Another important observation is that according to our calculations
the N concentration in Mn$_3X$N$_x$ ($X$=Rh, Ir, Pt) can be varied 
in the range between 0-20 at.\%,
thus providing an additional tuning parameter for the MAE, 
see Fig.~\ref{FIG:Mn3XvsN}. 
In all cases the interstitial alloys are on or close to the convex hull,
well below the $\Delta E_h <$ 50 meV/atom criterion for (meta-) stability.
However, a triangular spin structure is energetically preferred over
the collinear ferrimagnetic structure throughout the whole concentration range.
For Mn$_3$RhN$_x$, where the N 1b and 3d position 
lead to a gain in formation energy, more than 30 at.\% interstitial N
can be incorporated into the structure based on the $\Delta E_h$ criterion.
For these high N concentrations a ferrimagnetic structure with significant 
net magnetization around 2 $\mu_B$/f.u. and MAE around -1 meV/f.u.
is expected based on our calculations.
A partial occupation of the N 1b and 3d interstitial positions
may also explain deviations from an ideal
triangular spin structure reported in this compound~\cite{Fruchart78}.

\begin{figure}
\includegraphics[width=0.45\textwidth]{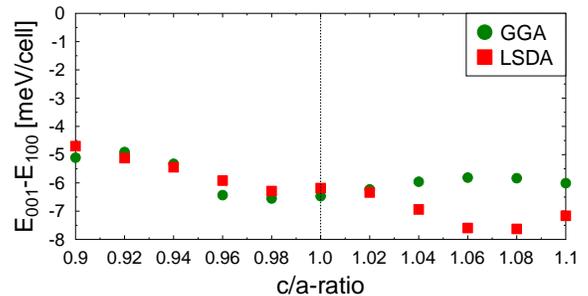}
\caption{\label{FIG:Epitax-Mn3IrN}Calculated MAE vs. $c/a$-ratio for Mn3IrN.
  Results are shown for GGA (filled green circles) and LSDA (filled red squares).
  The volume was fixed to the GGA optimized bulk value.}
\end{figure}

Fig.~\ref{FIG:Epitax-Mn3IrN} shows the calculated MAE as a function of $c/a$
for Mn$_3$IrN in the ferrimagnetic spin configuration employing 
selfconsistent calculations with an enhanced k-mesh. 
First, we note that the MAE obtained from selfconsistent calculations for the bulk value
is in good agreement with the MAE obtained with the force theorem in the high-throughput calculations,
underlining the suitability of the band anisotropy as a descriptor in
high-throughput calculations.
The results show that the huge MAE is robust against variations of the 
$c/a$-ratio in epitaxially grown films, with values between -4 and -8 meV/cell
for a wide range of lattice parameters.
In particular, the results show that the symmetry breaking due to the
ferrimagnetic spin arrangement is responsible for the huge MAE and not 
a slight distortion of the lattice from cubic symmetry ($c/a$=1).
The total spin moment ranges between 2 and 4 $\mu_B$/cell for $c/a$-ratios 
investigated. 
Thus, Mn$_3$IrN would be an excellent hard magnetic material if it could be stabilized in the ferrimagnetic spin configuration, which would outperform the
known hard magnet L1$_0$-CoPt with an MAE of about 0.5 meV/atom and a total magnetic moment of about 1.1 $\mu_B$/atom.
According to our VASP calculations, N has also a strong 
impact on the magnetic anisotropy 
in the NC ground state and enhances the energy difference $\Delta E= E_{\Gamma^{4g}}- E_{\Gamma^{5g}}$ between  
the easy $\Gamma^{4g}$ ground state and the $\Gamma^{5g}$ representation of the magnetic structure 
from about
-8 meV (Mn$_3$Ir) to -12 meV (Mn$_3$IrN). 
Based on these results we expect that interstitial N has also a strong 
impact on amorphous Mn$_3$Ir, which is one of the state-of-the-art
materials for providing exchange bias in hard magnetic films.
In case of Mn$_3$Pt(N), $\Delta E$ changes sign from -3 meV (Mn$_3$Pt) to
6 meV (Mn$_3$PtN), so that the $\Gamma^{5g}$ representation becomes the ground state.

According to our calculations, a relatively large epitaxial strain around 5\% 
is necessary to stabilize
a ferrimagnetic ground state in Mn$_3$IrN at 0 K, see Fig. 1 in the Supplementary. 
Surprisingly, the necessary strain
in Mn$_3$Ir is lower, despite the smaller energy difference between
the NC and ferrimagnetic state in Mn$_3$IrN, possibly indicating that an intermediate
N concentration is most suitable for a stabilization of the ferrrimagnetic
state.
Further, we note that in the related compound Mn$_3$GaN calculations by 
Zemen {\em et al.}~\cite{Zemen17}
have shown that a combination of strain and temperature leads to 
a stabilization of a ferrimagnetic ground state.
A similar behaviour was recently also experimentally observed for a Mn$_3$NiN film on a SrTiO$_3$ substrate~\cite{Boldrin19}.
In passing we note also that epitaxial strain in Mn$_3X$N as well as in Mn$_3X$ ($X$=Rh, Ir, Pt)
is accompanied by a significant net magnetization in the NC ground state, albeit smaller
than the one predicted for Mn$_3$SnN~\cite{Zemen17a}. This suggests that a piezomagnetic
effect may be observed in a wide range of NC antiferromagnets.
A more detailed analysis is beyond the scope of the present 
high-throughput study and left for future experimental and theoretical work.

\begin{figure}
\includegraphics[width=0.45\textwidth]{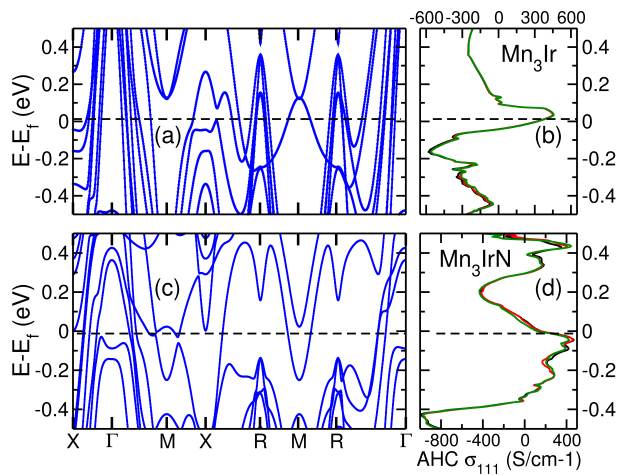}
\caption{\label{FIG:AHE}Calculated band structure and AHC of Mn$_3$Ir [(a) and (b)] and Mn$_3$IrN [(c) and (d)] respectively. Fig. (b) and (d) the black, red and green line correspond to the $\sigma_x$ and $\sigma_y$ and $\sigma_z$ AHC tensor.}
\end{figure}

The anomalous Hall conductivity (AHC) and anomalous Nernst coefficient (ANC) depend on the spin configuration of the system. 
The origin of non-vanishing AHC in $\Gamma_{4g}$ magnetic ordering has been studied based on symmetry analysis~\cite{Gurung2019,ilias-MnGaN}. 
Fig.~\ref{FIG:AHE} shows the calculated band structure and anomalous Hall conductivity (AHC) of Mn$_3$Ir and Mn$_3$IrN in the $\Gamma^{4g}$ ground state.
For both compounds. the calculated AHC $\sigma_{111}$ is significant, with 315 S/cm for Mn$_3$Ir and 157 S/cm for Mn$_3$IrN. 
This is in overall good agreement with a study by Chen {\em et al.}, where a moderately smaller AHC was calculated for Mn$_3$Ir (230 S/cm)~\cite{Chen14},
but differs in sign and magnitude from a recent AHC calculation by Huyen {\em et al.} for Mn$_3$IrN  (-573.3 S/cm)~\cite{huyen2019topology}.
The AHC has been shown to be highly sensitive to the choice of the magnetization direction~\cite{ilias-MnGaN}, which may be a reason for this discrepancy.
Based on our calculations, we predict a huge anomalous Nernst effect,
with calculated ANC $\alpha_{111}$ of 6195 and -4892 S/(cm eV) for Mn$_3$Ir and Mn$_3$IrN respectively. 
Noteworthy, there is change in sign of ANC for Mn$_3$IrN.
Both end members of the Mn$_3$IrN$_x$ (0$\leq x\leq$ 1) series show interesting transport properties. According to our thermodynamic calculations alloys with
intermediate N concentrations are stable, which may be a route for further tuning of the transport properties.

\section{Summary}
In summary we have shown that the interstitial elements N, C and B can have a strong impact on the magnetic properties of cubic compounds
and that interstitial alloying can be a fruitful approach to enhance the magnetic properties.
For reliable predictions of materials with enhanced properties, phase stability (i.e. calculation of the convex hull
including all possible decomposition reactions) is essential.
The high-throughput scheme introduced in this article is a unique computational tool to identify promising magnetic materials in order
to guide the design of new magnetic materials.

Based on our calculations, we expect ferromagnetic Ni$_3$MnN$_x$ (with about 11 at.\% N) and the ferrimagnetic compounds Mn$_3$SbN
and Mn$_3$RhN$_x$ (with about 30 at.\% N) to show bulk magnetic properties with sizable magnetization in the range of 0.5-1 $\mu_B$/atom
and sizable uniaxial MAE in the range of 100-200 $\mu$eV/atom.

The Mn$_3X$(N) ($X$=Rh, Ir, Pt and Sb) compounds are identified as materials with strong MAE and a sizable magnetization in a collinear ferrimagnet state. The calculations also show that --
with exception of Mn$_3$SbN --
these materials can not readily be used as bulk magnets, as their ground state is a non-collinear fully compensated antiferromagnet.
As a strong magnetic anisotropy is also present in the non-collinear ground state of these alloys, these materials are however of
strong interest for multilayer structures where the hard magnetic antiferromagnetic layer is exchange coupled to a soft magnetic
layer with large magnetization.
Especially for Mn$_3$Ir, which is already used as exchange bias material in hard magnetic films, interstitial alloying with N can be a promising approach to enhance the magnetic properties of a state-of-the-art material.
Further, an application of the strong MAE in the ferrimagnetic state may be possible in thin films,
but requires more detailed theoretical and experimental investigations. 
For Mn$_3$Ir and Mn$_3$IrN we find a significant anomalous Hall effect and a huge anomalous Nernst effect. 
Variation of the N concentration may be a route for further optimization of the topological transport properties,
as the Mn$_3$IrN$_x$ alloys share the same triangular spin arrangement.
Moreover, we have shown that interstitials can also have an impact on the ground state magnetic structure and can thus be used to tune magnetic transitions for instance in magnetocaloric applications. In Mn$_3$Sb(N) the magnetic ground state changes from NC to ferrimagnetic upon alloying with N. For Mn$_3$Sb, NC, ferrimagnetic and ferromagnetic magnetic states are very close in energy,
indicating that moderate fields may be sufficient to switch from a low to a high magnetization state.
This makes Mn$_3$Sb based alloys potentially interesting for magnetocaloric applications, but requires further studies.

For Mn alloys, our calculations show that the assumption of a ferromagnetic spin alignment, which is currently used
in high-throughput databases,  is unreliable to predict formation energies and phase stability.
The symmetry based methods applied in this article to determine the magnetic ground state can  pave the way
to predictions of the phase stability based on DFT-high-throughput calculations for these important alloy systems, which
are of interest e.g. for permanent magnet, magnetocaloric and spintronics applications.
\section*{Acknowledgements}
  We acknowledge fruitful discussions with Karl G. Sandeman and Zsolt Gercsi.
  Financial support by the DFG CRC/TRR 270, the German federal state of Hessen through its excellence programme LOEWE ''RESPONSE''
  and the European Community (NOVAMAG) is gratefully acknowledged.
The work of JZ was supported by the Ministry of Education, Youth and Sports of the Czech Republic from the  OP RDE programme under the project International Mobility of Researchers MSCA-IF at CTU No. CZ.02.2.69/0.0/0.0/18\_070/0010457, 
and from the Large Infrastructures for Research, Experimental Development and Innovations project IT4Innovations National Supercomputing Center-LM2015070.
Calculations for this research were conducted on the Lichtenberg high performance computer of the TU Darmstadt.

\end{document}